\documentclass[a4paper]{spie}  

\usepackage{amsmath,amsfonts,amssymb}
\usepackage{graphicx}
\usepackage[colorlinks=true, allcolors=blue]{hyperref}

\title{Towards First Light for MICADO: Validation of the PSF-R service with real data}

\author[a,b]{Matteo Simioni}
\author[a,c,b]{Luca Cortese}
\author[d]{Roland Wagner}
\author[a,b]{Carmelo Arcidiacono}
\author[a]{Andrea Grazian}
\author[a]{Marco Gullieuszik}
\author[e,b]{Elena Masciadri}
\author[f,b]{Elisa Portaluri}
\author[a]{Benedetta Vulcani}
\author[g]{Anita Zanella}
\author[h]{Richard Davies}
\author[i,j,k]{Johanna Hartke}
\author[l,b]{Roberto Piazzesi}
\author[m]{Stefan Raffetseder}
\author[l,b]{Piero Vaccari}
\affil[a]{INAF - Osservatorio Astronomico di Padova, Vicolo dell'Osservatorio 5, I-35122 Padova, Italy}
\affil[b]{ADaptive Optics National laboratory in Italy (ADONI)}
\affil[c]{Dipartimento di Fisica e Astronomia "G. Galilei", Università di Padova, Via Marzolo 8, I-35131 Padova, Italy}
\affil[d]{Johann Radon Institute for Computational and Applied Mathematics of the Austrian Academy of Sciences (RICAM), Altenberger Stra{\ss}e 69, A-4040 Linz, Austria}
\affil[e]{INAF - Osservatorio Astrofisico di Arcetri, Largo Enrico Fermi 5, I-50125 Florence, Italy}
\affil[f]{INAF - Osservatorio Astronomico d'Abruzzo, Via Mentore Maggini, I-64100 Teramo, Italy}
\affil[g]{INAF - Osservatorio di Astrofisica e Scienza dello Spazio di Bologna, Via Piero Gobetti 93/3, I-40129 Bologna, Italy}
\affil[h]{MPE - Max-Planck-Institut für extraterrestrische Physik Giessenbachstrasse 1, D-85748 Garching, Germany}
\affil[i]{Finnish Centre for Astronomy with ESO, (FINCA), University of Turku, FI-20014 Turku, Finland}
\affil[j]{Tuorla Observatory, Department of Physics and Astronomy, University of Turku, FI-20014 Turku, Finland}
\affil[k]{Turku Collegium for Science, Medicine and Technology (TCSMT), University of Turku, FI-20014 Turku, Finland}
\affil[l]{INAF - Osservatorio Astronomico di Roma, Via Frascati, 33, I-00078 Monteporzio Catone, Italy}
\affil[m]{Industrial Mathematics Institute, Johannes Kepler University (JKU), Altenberger Stra{\ss}e 69, A-4040 Linz, Austria}
\authorinfo{Further author information: send correspondence to M.S., e-mail: msimioni@inaf.it}

\begin{document} 
\maketitle

\begin{abstract}
MICADO, one of the ELT’s first-light instruments, will rely on point-spread-function reconstruction (PSF-R) to meet scientific requirements, especially for SCAO observations of targets that lack bright reference stars nearby. PSF-R will be performed in post-processing using only AO telemetry, recorded simultaneously with the science data. For SCAO observations, both on- and off-axis PSFs will be delivered. We present the current status and results from an ESO TTR campaign begun in 2023, targeting a sparse three-star asterism under varying conditions with the ERIS instrument at VLT. Reconstructed PSFs reach $\sim10\%$ accuracy in Strehl ratio and FWHM at separations up to half the size of the isoplanatic patch. The current results suggest the telemetry-driven approach can reproduce not only the core PSF structure but also the extended wings, which are essential for reliable photometry and morphological analysis of bright extended sources. The obtained results also corroborate the key role of wind as a limiting factor of the method, reducing the amount of recovered information on specific directions.
\end{abstract}

\section{INTRODUCTION}
\label{sec:intro}  
By 2030, the ESO Extremely Large Telescope (ELT) will be operational, equipped with the Multi-AO Imaging Camera for Deep Observations (MICADO \cite{Davies16}). Given the temporal and spatial variability of the PSF in such an AO-assisted instrument, accurately characterising the PSF plays a crucial role for exploiting the full scientific potential of the instrument: the final MICADO user will be provided with a PSF reconstruction (PSF-R) service\cite{Simioni20,Grazian22,Grazian24}.
In the case of MICADO single-Conjugate AO (SCAO) on-axis observations (i.e., observations of the AO reference source), a telemetry-only, PSF-R method has been proposed\cite{Wagner18}. In Ref.~\citenum{Wagner23}, the MICADO SCAO PSF-R method has been extended also to the off-axis case using, again, a telemetry-only approach that does not require any information coming from the scientific frames. This is possible thanks to a temporal tomography of the reconstructed incoming wavefront, assuming the knowledge of the atmospheric conditions and precise instrumental calibrations. The MICADO PSF-R has been tested in both ELT-like end-to-end (E2E) simulations\cite{Wagner23} and on real SCAO data obtained with LUCI@LBT\cite{Simioni22} and ERIS@VLT\cite{Simioni23,Simioni24} . Specifically, to validate the MICADO SCAO PSF-R method, an ESO Technical Time Request (TTR) programme is being carried out involving ERIS@VLT. In the present work, we will focus on the preliminary results obtained in the context of this ESO TTR programme, providing also a simulation framework to accurately contextualise these results with the application of the PSF-R method to MICADO SCAO.

\section{DATA}\label{sec:data}
\subsection{ERIS@VLT ESO TTR}\label{sec:real}
Starting from 2023, in the context of the ESO TTR, a few datasets have been collected so far. All refers to the same bright (H=$9 - 9.5$ VEGAmag), isolated asterism of 3 stars collected in different atmospheric conditions and filters. The 3 stars are the AO reference star, an off-axis source at $7\,$arcsec and another one at $16\,$arcsec\cite{Simioni23,Simioni24} .

\begin{table}[ht]
\caption{Log of the ERIS@VLT observations considered in this work. For each set, effective wavelength ($\lambda$) and MASS-DIMM isoplanatic angle ($\theta_0$) measurements are also included.} 
\label{tab:erislog}
\begin{center}       
\begin{tabular}{|l|c|c|c|c|} 
\hline
SET ID & DATE & NDITxDIT & $\lambda$ & $\theta_0$ \\
\hline
ERIS2023\_231915 & 2023-04-25T23:19:15.1853 & $225\times1\,$s & $1.65\,{\rm \mu m}$ & $4^{\prime\prime}.114$\\
\hline
ERIS2023\_233637 & 2023-04-25T23:29:07.2387 & $225\times1\,$s & $1.65\,{\rm \mu m}$ & $3^{\prime\prime}.86$\\
\hline
ERIS2023\_233637 & 2023-04-25T23:37:36.6705 & $225\times1\,$s & $1.65\,{\rm \mu m}$ & $3^{\prime\prime}.831$\\
\hline
ERIS2024\_234236 & 2024-03-26T23:42:47.4785 & $280\times0.9\,$s & $1.28\,{\rm \mu m}$ & $2^{\prime\prime}.051$\\
\hline
ERIS2024\_235036 & 2024-03-26T23:51:14.0755 & $280\times0.9\,$s & $1.65\,{\rm \mu m}$ & $2^{\prime\prime}.412$\\
\hline
ERIS2024\_235754 & 2024-03-26T23:58:45.2894 & $280\times0.9\,$s & $2.2\,{\rm \mu m}$ & $2^{\prime\prime}.67$\\
\hline
\end{tabular}
\end{center}
\end{table} 

The log of the full dataset considered in this work is presented in Table~\ref{tab:erislog}. Both in 2023 and 2024, the observations have been collected within $30$ minutes, with fairly constant atmospheric conditions over this time interval. While in 2023 all the data had been collected using the same narrow-band filter, in 2024 each set was acquired using different filters. The effective wavelength is reported in Table~\ref{tab:erislog}, for each case, along with the MASS-DIMM isoplanatic angle. 

\subsection{Simulations}\label{sec:simu}
To fully interpret the results, we produced a set of simulation of a SCAO, ERIS-like system. Both AO telemetry and corresponding focal plane data have been produced using SPECULA\cite{Rossi26}. A simple 3-layer atmosphere model has been considered, representing the first step towards a more realistic set of simulations.

For the reconstruction of the simulated data, we used the same recipe used for real data and also in this case, the evaluation of the result has been performed comparing reconstructed PSFs with simulated focal plane one. For the comparison, we used the on-axis and $12$ additional off-axis directions as a reference. The off-axis PSFs used in the evaluation are distributed as follows: $4$ at a radial distance of $7^{\prime\prime}$, $4$ at $30^{\prime\prime}$ and $4$ placed at $42^{\prime\prime}$. We limit ourselves to the case where the simulated focal plane observations are at $2.2\,{\rm \mu m}$.

\section{METHOD}\label{sec:method}
Our PSF-R approach does not require any information coming from the scientific frames (e.g., no PSF extraction from images). It relies on AO telemetry, atmospheric condition monitoring and accurate instrumental calibrations\cite{Wagner18,Wagner23,Simioni22} . In particular, for the SCAO off-axis PSF-R, turbulence and wind vertical profiles are necessary input for our PSF-R method\cite{Wagner23} .  
For the turbulence profile, MASS-DIMM measurements are used, retrieved from the ESO Ambient Condition Database. 

For the wind, we rely on predictions from the FATE\cite{Masciadri24} project (see also Ref.~\citenum{Masciadri13}). The wind profile is used by our tomographic (in time) method to set the motion of atmospheric layers. For this reason, it directly affects our reconstructions.

In all cases, the PSF-R has been performed compressing all the information of the atmospheric conditions into $3$ main layers. The height of each layer has been computed identifying the $3$ most relevant ones from the measured turbulence profile. For completeness, we note that, given the relative stability of atmospheric conditions, especially for all sets of 2023, turbulence profiles were rather similar. The wind vectors have been computed for each selected layer from the FATE predictions and kept constant between all sets pertaining to the same observing night. This will be discussed in more detail in the following. The same PSF-R recipe, calibrated for the ERIS instrument, has been used for all dataset, regardless of the atmospheric conditions or the used filter. 
Finally, it is worth noting that for the dataset of 2024, observations have been conducted at very low zenital distance ($<$5deg) wich produce a large rotation of the science field of view, but not of that of the wavefront detector. This effect has been taken into account in the analysis. 

\begin{table}[ht]
\caption{Parameter values used for the SPECULA ERIS-like simulation.} 
\label{tab:ssim}
\begin{center}       
\begin{tabular}{|l|c|c|c|} 
\hline
Parameter & Layer 1 & Layer 2 & Layer 3 \\
\hline
layer height & $68\,$m & $2955\,$m & $10370\,$m \\
\hline
turbulence fraction & $50\%$ & $25\%$ & $25\%$ \\
\hline
wind speed & $5\,$m/s & $10\,$m/s & $15\,$m/s \\
\hline
wind direction & $0\,$deg & $90\,$deg & $45\,$deg \\
\hline
\end{tabular}
\end{center}
\end{table} 
For the SPECULA simulation, we collected the used value for all the relevant parameters in Table~\ref{tab:ssim}. The height of the $3$ selected layers is consistent with the one used for real observations. The turbulence fraction of the lowest layer (ground) is still plausible, but somewhat lower than what is expected in typical Paranal conditions. This enhances the contribution of the intermediate and upper layers, for which we also assume a relatively high wind speed.

\section{Results and Discussion}
   \begin{figure} [ht]
   \begin{center}
   \begin{tabular}{c} 
   \includegraphics[width=.99\columnwidth]{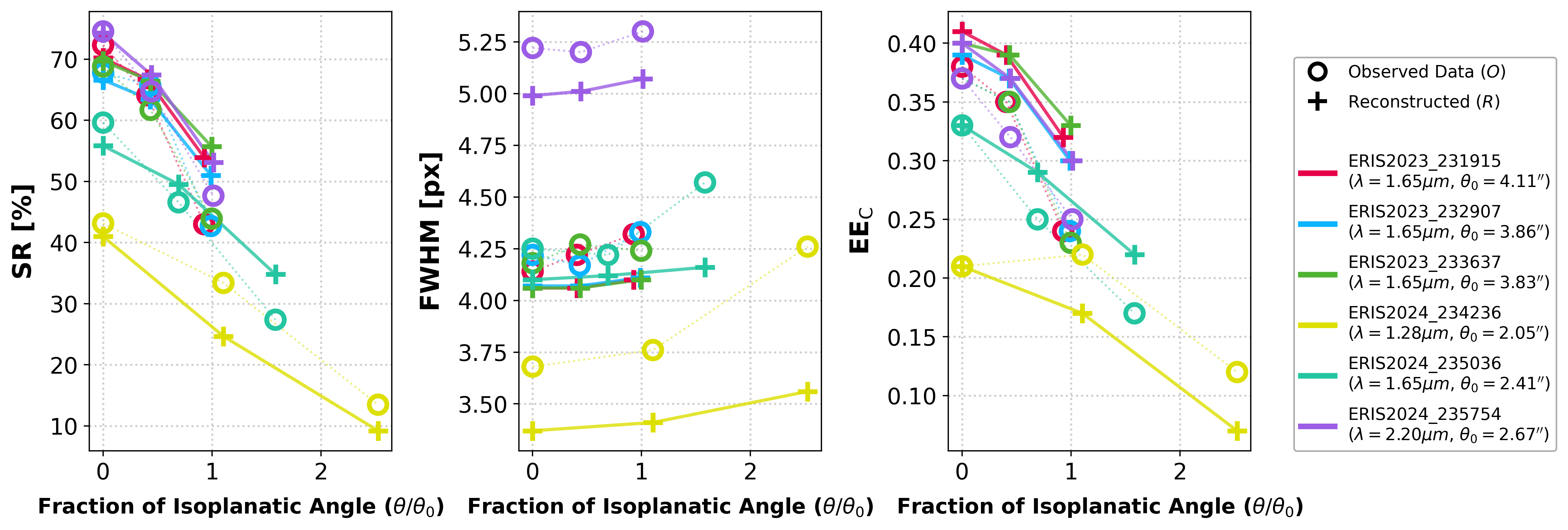}
   \end{tabular}
   \end{center}
   \caption
   { \label{fig:res_abs} 
Agreement between observed (circles) and reconstructed (crosses) PSFs for different ERIS@VLT datasets.  A classic PSF metric is provided. Different atmospheric conditions are represented by the isoplanatic angles (@500nm) reported in the legend ($\theta_0$). For each set, effective wavelength is also indicated in the legend ($\lambda$). In order to properly compare the results, we adopted the fraction of isoplanatic angle as a measure of the angular distance between the sources.}
   \end{figure}

In order to evaluate the performance of the reconstruction, we adopt a classic PSF metric \cite{Simioni23} composed by Strehl Ratio (SR), full-width at half-maximum (FWHM) and encircled energy in the PSF core (EE$_{\rm C}$). In Figure~\ref{fig:res_abs} we show, for each considered set of ERIS data, the measured absolute values of each of these PSF indices (open circles). In the same plot, we also show the reconstructed values (crosses). It can be noted the good agreement between observed and reconstructed PSFs. In particular, performances are comparable among all sets at fixed fractions of the isoplanatic angle. We used this quantity in order to properly compare the results in a reference frame independent of the specific atmospheric conditions or used filter.

\begin{figure} [ht]
   \begin{center}
   \begin{tabular}{c} 
   \includegraphics[width=.90\columnwidth]{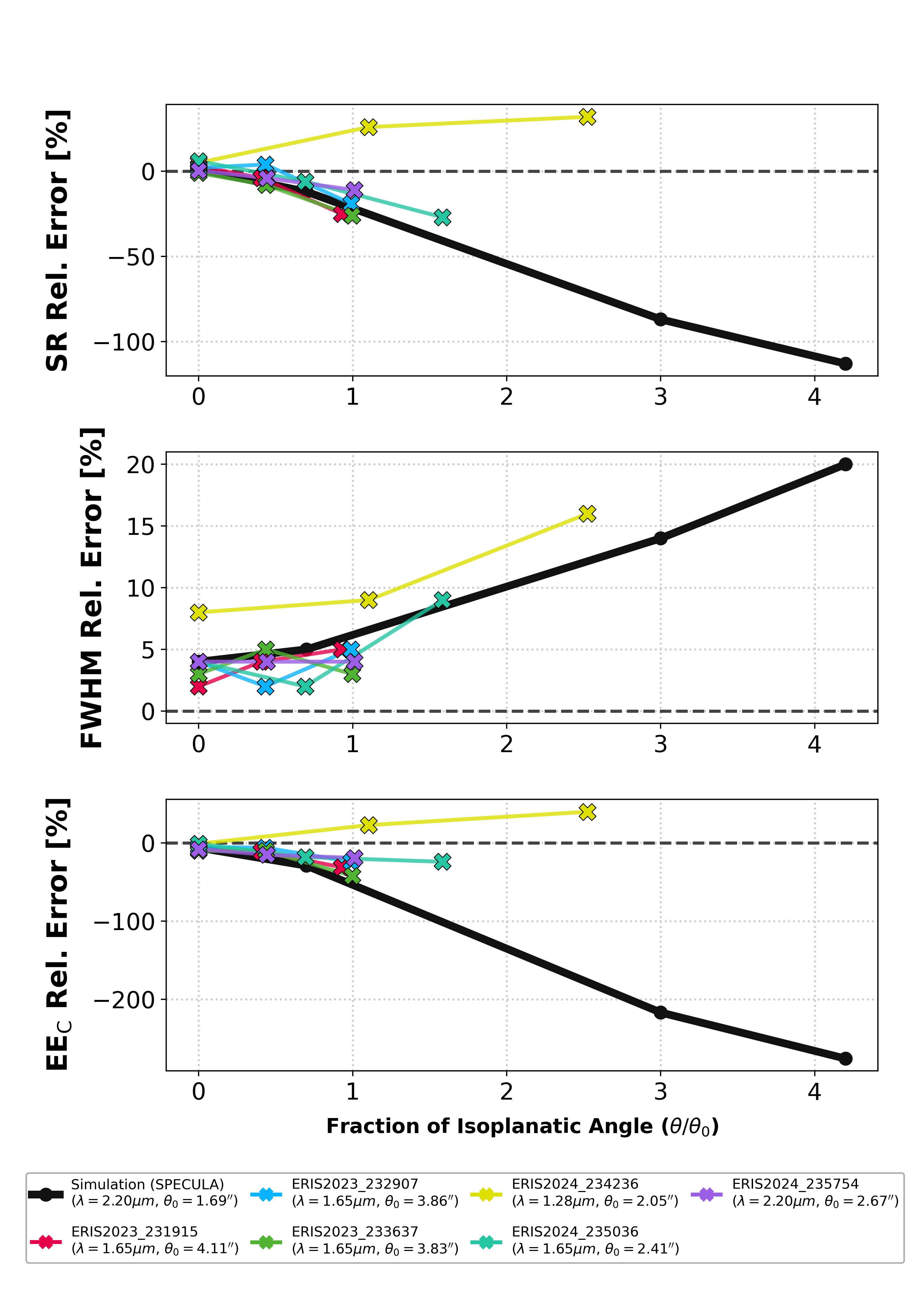}
   \end{tabular}
   \end{center}
   \caption
   { \label{fig:res_rerr} 
Performance of our PSF-R method expressed in terms of relative error on SR, FWHM and EE$_{\rm C}$. The black curve refers to the SPECULA simulation. For ERIS@VLT observations, the colour code is the same as in Fig.~\ref{fig:res_abs}. }
   \end{figure}

The performance of our SCAO PSF-R method is quantified in Figure~\ref{fig:res_rerr}, in terms of relative error on SR, FWHM and EE$_{\rm C}$. Each dataset is colour-coded as in Figure~\ref{fig:res_abs} and they are all consistent among each others. Moreover, the results from the SPECULA simulation nicely match the observed trend of relative errors as a function of off-axis distance for all the indices of the used PFS metric. It results that we are able to reconstruct the ERIS PSFs with an error of the order of $10\%$, or lower, in SR, FWHM and $EE_{
rm C}$ for off-axis distances up to half of the isoplanatic patch size. The relative error grows to about $20\%$ at $1$ isoplanatic angle. 

The only exception is the ERIS2024\_234236 set. In this case, it appears that an overestimation of the turbulence is present, resulting in larger SR and FWHM of the reconstructed PSF with respect to the observed ones. Consistently, EE$_{\rm C}$ reconstructed values are lower than the corresponding observed ones. A possible explanation for this behaviour is related to the fact that this set is also the one with the shortest wavelength. As already mentioned, we used the same PSF-R recipe for all the sets. This prevents any wavelength-dependent fine-tuning for the reconstructed PSFs, which may become relevant in this case. Further investigations are required to confirm this possibility.

As the effect of errors on wind vectors is concerned, in reconstructing the simulated data, we assumed a perfect knowledge of the true input values. The same assumptions have been made for the turbulence profile. However, the present results demonstrate that the approximate input values used in the case of real data already produce the expected level of precision. This suggests that, in the case of an 8m-class telescope, the accuracy of the FATE predictions is sufficient to obtain high precision, telemetry-only, reconstructed PSFs in the case of SCAO observations. The contribution of uncertainty on wind vectors, even if a second-order effect, will be evaluated in a dedicated study. 

Comparing the present results with E2E MICADO simulations\cite{Wagner23,Simioni23,Simioni24} suggests that in the case of ERIS observations, our PSF-R method is underperforming. This can be related to the physical size of the pupil projection of the ELT is bigger than that of the VLT. Sampling a larger patch of the sky, MICADO will be able to provide more information to compute the temporal tomography from the telemetry of its SCAO module. This increase in spatial information of the incoming wavefront translates into better PSF-R performances. 
\clearpage

\section{CONCLUSION}\label{sec:fin}
The MICADO telemetry-only PSF-R\cite{Wagner23} enables the derivation of accurate SCAO off-axis PSF templates. 
The method is currently being validated on real SCAO data coming from 8m-class telescopes with extremely encouraging results, reaching precision of the order of $10\%$ or less in SR, FWHM and EE$_{\rm C}$ up to off-axis distances of half the size of the isoplanatic patch.
Simulations of ERIS-like systems corroborate these results and are key to linking them to MICADO ones.
Comparing the results obtained with ERIS@VLT data with end-to-end MICADO simulations suggests that the increase in pupil size (from 8-m to 39-m class) will enhance PSF-R performance, meaning the telemetry-only, SCAO PSF reconstruction is expected to be better for MICADO.

\acknowledgments 
The Italian authors warmly thanks STILES for the generous support to the MICADO activities. STILES - STrengthening the Italian Leadership in ELT and SKA is a program funded by the National Recovery and Resilience Plan (PNRR, Mission 4 Component 2 Investment 3.1 Project STILES IR0000034 – CUP C33C22000640006) which aims to strengthen Italian leadership in the exploration of the Universe by developing laboratories and instruments for the two largest ground-based telescopes of the coming decades: the Extremely Large Telescope (ELT) and the Square Kilometer Array (SKA). STILES is a program coordinated by the National Institute for Astrophysics (INAF) in which 7 Italian universities participate, and it is carried out in collaboration with international research institutes. The project began in 2023 and officially ended in April 2026.

\bibliography{report} 
\bibliographystyle{spiebib} 

\end{document}